\documentclass{svmult}

\usepackage{mathptmx}
\usepackage{amsmath,amssymb}
\usepackage{helvet}
\usepackage{courier}
\usepackage{multicol}
\usepackage{float}
\usepackage{geometry}
\usepackage{graphicx}
\usepackage[english]{babel}
\usepackage{natbib}

\begin{document}

\title*{Planetesimals and satellitesimals: formation of the satellite systems}
\titlerunning{Formation of satellite systems}
\author{Ignacio Mosqueira \and Paul Estrada \and Diego Turrini}
\institute{Ignacio Mosqueira \at SETI Institute, 515 N Whisman Rd, Mountain View, CA 94043, United States, \email{imosqueira@seti.org}}

\abstract{
The origin of the regular satellites ties directly to planetary formation in that the satellites form in gas and dust disks around the giant planets and may be viewed as mini-solar systems, involving a number of closely related underlying physical processes. The regular satellites of Jupiter and Saturn share a number of remarkable similarities that taken together make a compelling case for a deep-seated order and structure governing their origin. Furthermore, the similarities in the mass ratio of the largest satellites to their primaries, the specific angular momenta, and the bulk compositions of the two satellite systems are significant and in need of explanation. Yet, the differences are also striking. We advance a common framework for the origin of the regular satellites of Jupiter and Saturn and discuss the accretion of satellites in gaseous, circumplanetary disks. Following giant planet formation, planetesimals in the planet's feeding zone undergo a brief period of intense collisional grinding. Mass delivery to the circumplanetary disk via ablation of planetesimal fragments has implications for a host of satellite observations, tying the history of planetesimals to that of satellitesimals and ultimately that of the satellites themselves. By contrast, irregular satellites are objects captured during the final stages of planetary formation or the early evolution of the Solar System; their distinct origin is reflected in their physical properties, which has implications for the subsequent evolution of the satellites systems. 
}

\maketitle

The satellites of Jupiter and Saturn have received renewed interest as a consequence of the Cassini mission to Saturn, which has resulted in a growing body of literature concerning both irregular (see \citet{jah07} for a review) and regular satellites (see \citet{eea09} for a review; and \citet{caw09} for a different take). We make no claim to completeness -- the aim of this submission is not to provide a comprehensive review of the field, which has already been attempted in those publications, but to describe key physical processes and observational constraints. While there is unambiguous observational evidence that the regular and irregular satellites are distinct populations, there exist at least indirect ties between them when it comes to the processes involved in the capture of objects from heliocentric to circumplanetary orbits. There is also a hint of a direct connection in terms of the locations of the innermost irregular satellites and the size of the disk formed by gas inflow through a gap ($\sim R_H/5$; \citet{mae03a}; \citet{eea09}; \citet{aab09}); irregular satellites captured into orbits with smaller semi-major axes may be removed by gas drag from the outer disk and possibly contribute mass to the regular satellites, though this link remains to be quantitatively explored.\\
It is also worth noting that the processes shaping the formation of satellite systems are strongly coupled to those governing planetary formation and the evolution of the early Solar System. In fact, planetary formation sets the initial conditions for satellite formation which, therefore, cannot be treated in isolation. The forming satellite systems can exchange material with the outer Solar System, receiving solids mainly in the form of planetesimals captured or ablated due to the interaction with the circumplanetary gas disks, and even scattering satellitesimals out into heliocentric orbits. Moreover, the regular satellites, once formed, are still subject to exchanges with the outer Solar System, as dramatically illustrated by the disruptive capture of Triton in Neptune's satellite system. Below we discuss the formation and properties of both regular and irregular satellites in the framework of such exchange processes.

\section{Formation of the regular satellites of giant planets}

In the core accretion model of planet formation \citep{bap86} a core must first form by accretion of planetesimals. In this mode of planet formation most of the mass of solids is taken to reside in planetesimals of size $\sim 10$ km \citep{was93}. Planetesimals may dissolve in the envelopes of the forming giant planets, thus enhancing the planet's metallicity \citep{pea86}. Most of this high-Z mass delivery takes place before the cross-over time, when the mass of the gaseous envelope grows larger than that of the core. This stage is then followed by a dilution during runaway accretion, depending on the amount of gas accreted during this brief phase of the planet's growth \citep{pea96}.
As the giant planet reaches its final mass, planetesimals in its feeding zone undergo collisional grinding. In the Jupiter-Saturn region, the collisional timescale for kilometer-sized objects is similar to the ejection timescale $\sim 10^4$ yrs, so that a fraction of the mass of solids will be fragmented into objects smaller than 1 km \citep{saw01,cam03}. The collisional cascade facilitates planetesimal delivery to the circumplanetary disk (in $\sim 10^4$ yrs) because smaller planetesimals are easier to capture.

\subsection{Satellite migration and growth}

\citet{mae03a} start with a disk populated by satellite embryos ($\sim 10^3$ km) and satellitesimals. These authors adopt a bimodal size distribution or two-groups method, as has been done in a number of publications that treat protoplanetary growth by the binary accretion of planetesimals (e.g., \citet{pea96,gls04}). There are close parallels, but some differences as well. In both the planet and satellite cases, the migration of the embryos
is dominated by the tidal interaction with the gas disk, whereas planetesimals and satellitesimals migrate due to aerodynamic gas drag. But in the satellite case the transition from one type of migration to the other overlaps and leaves no size unaffected, i.e., all object sizes from kilometers to full-sized satellites undergo migration. Another difference involves the capture cross-section of migrating planetesimals versus satellitesimals. Satellite embryos have a larger ratio of physical radius to Hill radius ($\sim 0.1$ at the location of Ganymede) than do protoplanets ($\sim 10^{-3}$ at the location of Jupiter). As a result, the impact probability of passing objects is larger (cf. fig. 9 from \citet{klg93}). 
Also, the resulting gravitational focusing factor ($F_g \sim 10$) for satellites is never as large as it is for planet embryos during their runaway phase, before they grow large enough to ``heat'' up their own food and their growth switches to oligarchic accretion\footnote{At this stage, satellitesimals should be interpreted as those objects which determine embryo growth, containing a significant fraction of the mass of solids in the disk (though other sizes may also contribute to the mass distribution).}.

\subsubsection{Timescale for disk formation and cooling}

While the gas accretion rate is high the planet's envelope takes up a fraction of the planet's Roche lobe. Thus, there is agreement in the literature that the formation of icy satellites close-in to the planet must await for gas accretion onto the planet to wane (\citet{cea89}; \citet{mea99}; \citet{aab09}). The timescale for Jupiter and Saturn to clear a gap in between ($\tau \sim 10^4$ years or less in a compact initial configuration) sets the timescale for the end of gas accretion\footnote{Provided the protoplanetary disk at the location of Jupiter and Saturn is weakly turbulent \citep{bea99}. In that case the gas inflow through the gap ceases.}. The timescale for envelope collapse is given by the Kelvin-Helmholtz time of $\sim 10^4$ years \citep{hbl05}, following which a circumplanetary disk will form. We stress that even after envelope collapse continued inflow onto the disk may keep it too hot for the concurrent accretion of close-in, icy satellites (see \citet{eea09} and \citet{aab09} for further details). Once the circumplanetary disk forms and gas inflow dies down (as the giant planets open a combined gap), the timescale for the disk to cool sufficiently for ice to condense depends on the disk opacity. The Rosseland mean thermal opacity for the gas is $K_{gas} \sim 10^{-4}$ cm$^2$g$^{-1}$, while that of micron-sized dust grains can be of order $\sim 1$ cm$^2$ g$^{-1}$ \citep{las82,cea01}. As particles grow, the opacity of the nebula decreases. A sustained high opacity may keep the subnebula weakly turbulent and hotter. Conversely, rapid coagulation allows the subnebula to become optically thin, cooling quickly and allowing turbulence to subside. Indeed, while the incremental growth of grains and dust may take place irrespective of the level of nebula turbulence, growth past the decoupling size likely requires very low levels of turbulence ($\alpha < 10^{-6}$) \citep{yas02,caw06}, which allows for a cool subnebula where condensation of volatiles takes place \citep{mae03a}. Following envelope collapse, ablation of meter-to-kilometer planetesimal disk crossers (see \citet{eea09} and references therein) and other mechanisms enhanced the concentration of solids in the satellite disk from a starting condition with condensable content enhanced by a factor of $3-4$ from cosmic mixtures, as suggested by the high-Z content of giant planet envelopes \citep{aea99}. Note that the circumplanetary disk and the giant planet envelope are both enriched in solids as a result of a common mechanism, i.e., planetesimal ablation (see Fig. 1 upper left panel; \citet{mec09}); however, the quantitative degree of enrichment may differ substantially between the two\footnote{The runaway gas accretion phase is expected to result in a dilution of the solid content of the planet's envelope, as most of the mass of solids is in planetesimals that are not coupled to the gas (see \citet{eea09} for a detailed discussion).}. Given a density of $\sim 1$ g cm$^{-3}$, a planetesimal of size $< 1$ km encounters a gas column equal to its mass when crossing the circumplanetary gas disk. Such a population of objects is likely to result from the fragmentation cascade following giant planet formation (e.g., \citet{cam03}). Thus, we consider a solids-enhanced, quiescent disk with optical depth $\tau \gtrsim 1$ and gas surface density $\Sigma_g \sim 10^4$ g/cm$^2$. The disk cooling time is then given by $\tau_{KH} \approx \Sigma_g C T_c/(2 \sigma_{SB} T_d^4) \sim 10^2 (250 K/T_c)^3 \tau$ yrs, where $\sigma_{SB}$ is the Stephan-Boltzmann constant, $C$ is the specific heat, and $T_c$ and $T_d$ are the midplane and photospheric temperatures.

\subsubsection{Disk scales}

Gas flowing into the protoplanet's Hill sphere forms a circumplanetary disk. One can obtain an estimate of the expected disk size by assuming that in-falling gas elements conserve specific angular momentum. Assuming that {\it prior} to gap-opening the giant planet accretes gas with semi-major axis originating within $\sim R_H$ from its location, centrifugal balance yields a characteristic disk size of $\sim R_H/50$ \citep{cap76,sea86,mae03a}. For Jupiter and Saturn these radii are located close to the positions of Ganymede and Titan, respectively. On the other hand, {\it after} gap-opening accretion continues through the planetary Lagrange points. In this case, the estimated characteristic disk size formed by the inflow is significantly larger, of the order of $\sim R_H/3$ (\citet{mae03a}, but note that this paper settles on an outer disk size of $\sim R_H/5$, based on the locations of irregular satellites).
In the absence of a gap, 3D effects allow low angular momentum gas to be accreted directly onto the planet, or in a compact disk of characteristic size $< 0.1 R_H$ \citep{akh03}, whereas in the presence of a well-formed gap a 2D treatment applies and the resulting disk size will be $> 0.1 R_H$. \citet{akh03} conclude that, for nominal disk parameters, two dimensional computations are reliable for secondary to primary mass ratio $\mu \gtrsim 10^{-4}$, such that the Hill radius of the protoplanet is larger than the disk scale-height. \citet{eea09} and \citet{aab09} show that the specific angular momentum of the inflow through the gap is about a factor of $3-4$ larger than that of the Galilean satellites $1.1 \times 10^{17}$ cm$^2$/s, which means that the gas will achieve centrifugal balance at a radial location of $\sim 100 R_J$. This disk is compact compared to the Hill radius, but it is
extended in terms of the locations of the regular satellites, and possibly linked to the location of the irregular satellites.

\subsubsection{Model parameters}

\citet{mae03a} divide the circumplanetary disk into inner and outer regions. For Jupiter, they compute the solids-enhanced minimum mass (SEMM) gas densities in the inner and outer disks based on the solid mass required to form Io, Europa (both re-constituted for missing volatiles) and Ganymede in the inner disk and Callisto in the outer disk. Inside of the centrifugal radius the surface gas density $\Sigma_g$ is $10^4-10^5$ g cm$^{-2}$ (corresponding to pressures $\sim 0.1$ bar), which yields an optical depth due to absorption by hydrogen molecules (assuming dust coagulation) $\tau_v \sim \Sigma K_{gas} \sim 1$. Outside the centrifugal radius $r_c \sim 30 R_J$, the gas surface densities are in the range $10^2-10^3$ g cm$^{-2}$, which results in a low optical depth in the range $\tau_v \sim 0.01 - 0.1$. These authors apply the same procedure to the circumplanetary disk
of Saturn by employing the masses of Titan and Iapetus to set the inner and outer disk masses, respectively, and choose a simple model where the gas density follows a simple $1/a$ dependence inside of $r_a \sim 20 R_J$ and outside of $r_b \sim 26 R_J$. The transition region has a width of (at least) $\sim 2H_c$, where $H_c$ is the subnebula scale-height at the centrifugal radius. This choice ensures that the gradient in gas density is not so steep as to lead to a Rayleigh-Taylor instability (e.g. \citet{lap93}). The use of a two component subnebula hinges on the assumption of decaying turbulence as Roche-lobe gas inflow ebbs. But it should be noted that detailed simulations of the formation of such a disk remain to be carried out\footnote{Here we rely mostly on the overall gas surface density of the subnebula, which we justify on observational grounds.}. 
We tie the disk's temperature profile to the planetary luminosity
at the tail end of giant planet formation \citep{hbl05}. For now, we use a heuristic temperature profile of the form $T = 3600 R_J/a$ for Jupiter (e.g., \citet{las82}) and $T = 2000 R_S/a$ for Saturn. The outer disks of Jupiter and Saturn have roughly constant temperatures in the range of $70 - 130$ K for Jupiter and $40 - 90$ K for Saturn, depending on solar nebula parameters.\\
{\it Inner disk:} In analogy to planetary accretion, we begin by calculating characteristic object sizes in such a disk. The characteristic size over which the self-gravity of the solids overcomes the shear is $l_1 = r_h$, where $l_1 = (m_1/\pi \Sigma_s)^{1/2}$, $\Sigma_s \sim 10^3$ g/cm$^2$ is the surface density of solids, and $r_h = a (m_1/3 M_P)^{1/3}$ is the Hill radius of a satellitesimal of mass $m_1$. The second characteristic mass can be obtained from the condition $l_2 = r_H$, where $l_2 = m_2/(4 \pi a \Sigma_s)$ is the distance over which the mass $m_2$ in a disk annulus with semimajor axis $a$ can force close encounters with passing satellitesimals, and $r_H = a (m_2/3 M_P)^{1/3}$ is the Hill radius of an embryo of mass $m_2$. For object densities of $\rho_s \sim 1$ g/cm$^3$, these characteristic masses correspond to sizes of $r_1 \sim 1 - 10$ km and $r_2 \sim 1000$ km. We now consider a gaseous disk with surface density $\Sigma_g \sim 10^4$ g/cm$^2$ and aspect ratio $H/a \sim 0.1$ (or a temperature of $\sim 250$ K at Ganymede [$15 R_J$], and $\sim 100$ K at Titan [$20 R_s$]), and obtain a Toomre parameter $Q \sim 100$, which shows that self-gravity is unimportant. The particle decoupling size is $\Omega t_s = 16 \rho_s r_p v_k/(3 C_D \Sigma_g c \Omega) \sim 1 \rightarrow r_p \sim 10$ m, where $t_s$ is the stopping time, $v_k$ is the Keplerian velocity, $C_D \sim 1$ is the drag coefficient\footnote{There is an issue whether this drag coefficient should include a correction in cases where the satellite can gravitationally perturb the gas flow (e.g., \citet{caw02}). \citet{ost99} argues that this enhanced drag is already incorporated into the migration rate due to differential tidal torques.} (see e.g. \citet{ahn76}), and $c$ is the sound speed. The gas drag time for a satellitesimal of size $r_1$ located at $15 R_J$ is $\tau_{drag} = 1/2 (v_k/c)^2 t_s \sim 100$ years. The gas drag time for an embryo of size $r_2$ is $\sim 10^5$ years. Such an embryo will also migrate due to the tidal interaction with the gas disk in a similar timescale $\tau_{I} \approx C_I \Omega a^2 c^2/(\Sigma_g G^2 m_2) \sim 10^5$ years, where $G$ is the gravitational constant and $C_I \lesssim 1$ in a 3D disk \citep{ttw02,akh03,bea03}. Thus, the crossover size between the regimes of gas-drag and tidal-torque
migration is $\sim 500-1000$ km. The Type I migration timescale for Ganymede is $\sim 10^4$ years, which is comparable to its formation time given by the gas drag time of satellitesimals of size $\sim 100$ km.\\
The question arises whether the satellite growth times are faster than the migration times. Ignoring gravitational focusing, the Safronov binary accretion timescale for embryos is $\tau_{acc} \sim \rho_s r_2/(\Sigma_s \Omega) \sim 10^3$ years, which is faster than either the Type I or gas drag migration for such objects in the SEMM disk. But note that if all the satellitesimals resided in objects of size $r_1 \sim 1$ km, then the time to clean the disk by gas drag migration would be $\sim 100$ years. The mass growth timescale due to dust and rubble sweep up as a result of differential drift is $\tau_d = 4 \rho_s r_2/(3 \rho_p \Delta v_p) \sim \rho_s r_2 a/(\Sigma_g \Omega) (a/H) \sim 10^3$  years, where $\Delta v_p \sim \eta v_k$ and $\eta = (v_k - v_g)/v_k$ is the fractional difference between the local Keplerian velocity $v_k$ and the gas velocity $v_g$ due to gas pressure support. This estimate assumes that the density of the dust and rubble layer settles to a state such that $\rho_p \approx \rho_g$,
where $\rho_g$ is the nebular gas density, which is consistent with a quiescent disk stirred only by weak, local production of turbulence at the shear layer. Thus, satellitesimals grow as they drift: We interpret Hyperion (whose radius is $135 \pm 4$ km, \citet{tea07}
) as a satellitesimal that was captured into resonance by gas drag inward migration (see section \ref{hyperion}). For objects of this ($\sim 100$ km) size, satellitesimal migration, and satellite migration and growth all take place on a similar timescale. This is because satellite growth is regulated by the rate at which the embryo's feeding zone is replenished by satellitesimals or embryos. Also, the Type I migration timescale of satellites is comparable to the gas drag migration timescales of such satellitesimals (cf. Titan-Hyperion).\\
{\it Outer disk:} Further out the characteristic sizes are smaller despite the increase in semi-major axis. The reason is that the surface density is taken to drop outside the centrifugal radius. Typical sizes in the outer disk for $\Sigma_s \sim 10$ g/cm$^2$ are $r_1 \sim 1$ km and $r_2 \sim 100$ km. The Safronov binary accretion time to form an embryo of size $r_2$ in the outer disk is $\sim 10^5-10^6$ years. The time
for gas drag to clear the outer disk of such embryos is also $10^5-10^6$ years, which \citet{mae03a} tie to Callisto's formation time.

\subsubsection{Satellite survival}

So far we have implicitly assumed that the relatively massive gas component of the circumplanetary disk is static and unaffected by the evolution of the solids. However, the SEMM model of \citet{mae03b} relies on gap-opening for satellite survival, which means that at some point during their migration and growth satellites become sufficiently massive to open a gap in the gas. There are uncertainties involved in estimating the value for the gap-opening masses, such as the role of 3D effects and stratification, but they are unlikely to greatly affect the inviscid gap-opening criterion of \citet{raf02} (see also \citet{lea09}), which is based on damping of 2-D acoustic waves by wave steepening \citep{gar01}.
For Jupiter and Saturn Rafikov's criterion yields gap-opening masses at $15 R_J$ and $20 R_S$ in disks with $h \simeq 0.1$ equal to the masses of Ganymede and Titan for disk surface densities of $\sim 2 \times 10^4$ g/cm$^2$ and $ 1 \times 10^4$ g/cm$^2$, respectively. It should be noted that the gap-opening mass depends on the semi-major axis $a$. Also, we assume a surface density profile $\Sigma_g \propto 1/a$, and an extended giant planet following envelope collapse with radius $R_P \sim 2 R_J$ (e.g., \citet{hbl05}), where $R_J$ is Jupiter's radius. As a satellite migrates inwards, it finds it easier to open a gap for a given surface density, but the surface density is taken to increase closer to the planet. While the timescale for accretion decreases closer-in, the satellite's feeding zone becomes smaller, and its inward migration speeds-up. Also, both the migration rate and the gap-opening condition are sensitive to the aspect ratio $H/a$.

\subsection{Observational constraints}

\subsubsection{Capture of Hyperion into a 4:3 resonance with Titan}\label{hyperion}

The origin of Hyperion in the $4:3$ mean-motion resonance with Titan presents a significant challenge. A tidal origin of resonance capture as may apply to Galilean satellites \citep{sam97} does not apply to the case of Titan and Hyperion. Given Titan's size and distance from Saturn, significant expansion of its orbit would require Saturn's dissipation parameter $Q$ to be much lower than the lower limit set by the proximity of Mimas \citep{lap00}. Work by \citet{wad85} showed that the combination of gas drag and perturbations due to mean motion resonances (MMR's) with a planet can have important consequences for the orbital evolution of small planetesimals. \citet{klg93} obtained a critical value for the drag parameter at which resonance trapping by a protoplanet breaks down. \citet{mal93} points out that resonance trapping is vulnerable to planetesimal interactions. This issue also arises in the satellite case, as the capture of Hyperion in a $4:3$ MMR with Titan illustrates.

\subsubsection{Callisto's internal state}

A challenging result is that the Galileo mission moment of inertia data are consistent with a fully differentiated Ganymede, but only a partially differentiated Callisto \citep{aea01}. While it is possible that non-hydrostatic effects in Callisto's core could be large enough to allow for complete differentiation of Callisto (see e.g. \citet{mck97,sea03}), its internal state may be a result of its accretion history. In this view, Callisto forms by accreting the volatile-rich condensables present in the extended, low density outer disk; its long formation timescale is tied to the disk clearing time, which is the time it takes for gas drag to clear the circumplanetary disk of solids \citep{mea01,mae03a,amb05}. That is, accretion in the sparse, extended outer disk takes a sufficiently long time that the heat of accretion can escape, and the satellite may not differentiate fully. Callisto could then be said to be the result of slowly
assembling thousands of volatile rich, ``cold'' embryos. Thus, the issue of Callisto's state of differentiation is linked to the observation of the empty space outside Callisto and inside the irregulars. A related question can be posed for Saturn: namely, why is Iapetus stranded far from the planet? In each case, it is natural to expect that gas drag clearing of swaths of the circumplanetary disk is involved in the answer\footnote{Note that the capture of Hyperion into resonance also involves gas drag inward migration.}. Iapetus' separation from Titan provides strong indirect support for a two-component subnebula: a dense inner region roughly out to the centrifugal radius near Titan, and a lower surface density tail extending out to as far as irregular satellite Phoebe. Similarly, Callisto derives its mass from a more extended region than does Ganymede, which can explain its partially differentiated state without resorting to fine-tuning poorly known parameters; that is, in a two component subnebula it is natural to expect both that a) some satellites form in the outer disk and others in the inner disk, and b) that satellites forming in the outer disk take significantly longer to accrete than those forming in the inner disk.\\
The issue arises whether typical impactors would bury heat in Callisto so deeply that it would lead to full differentiation of this satellite \citep{mck06}.
\citet{mae03a} estimate that outer disk satellitesimals could grow as large as several hundred kilometers as they migrate inwards due to gas drag. These sizes are computed by assuming that all collisions among satellitesimals are accretionary. However, as satellitesimals approach a satellite's feeding zone they are excited to velocity dispersions well in excess of their escape speeds, which is likely to result in non-accretionary, disruptive collisions (as Hyperion's irregular shape may attest to). Thus, it is reasonable to expect that Hyperion-like or smaller satellitesimals are typical impactors during the late-stage accretion of both Callisto and Titan.

\subsubsection{Compositional constraints}

Regular satellites may provide a probe of the compositional and thermal state in the subnebula at the tail end of giant planet formation. Early models propose the formation of satellites in the circumplanetary nebulae of Jupiter and Saturn \citep{par74,pea76}. These models envision a condensation sequence analogous to that of the nebula. Still, the physical conditions in the circumplanetary disks are thought to be sufficiently
different from those of the nebula that a sharp contrast has been drawn between objects forming in the outer nebula and giant planet subnebulae \citep{paf81}.
Yet, such ideas retain a phenomenological character, as the nature of the interaction between the two environments remains poorly understood. The compositional gradient of the Galilean satellites may provide a link to the environment in which they form (see \citet{eea09} and references therein). The similarities in the bulk properties of the regular satellites of Jupiter and Saturn strongly favour a unified framework for their origin; yet, the inner, icy satellites of Saturn exhibit no clear compositional trend (and may have been collisionally disrupted\footnote{Indeed, a number of factors, such as the stochastic compositional gradient of the inner Saturnian satellites, the majestic icy rings, and Titan's isolation and eccentricity, argue in favour of collisional processes (see \citet{mae05} for an intriguing, though intricate, possibility).}). In fact, recent Cassini results make a strong case that the medium-sized Saturnian satellites have densities \citep{jea06} that preclude
a solar composition ice/rock ratio (e.g., \citet{jal05} and references therein; \citet{mae05}). On the other hand, a persistently hot subnebula can prevent the condensation of volatiles close to Jupiter \citep{par74,las82}. Thus, the thermal conditions under which coagulation and growth take place may hold a key to the formation of the regular satellites.\\
{\it Compositional Constraints of the Outer Regular Satellites:} \citet{mec09} focus on the large, outer regular satellites of each satellite system: Ganymede and Callisto in the case of Jupiter, and Titan and Iapetus for Saturn. For objects the size of Iapetus or larger, the porosity is likely to be small not only because the internal pressure is large enough to close pore spaces, but also because the presence of short-lived radioactive nuclides heats the interior causing ice to flow. For such large satellites, densities can be interpreted in terms of rock/ice fractions. Iapetus' low density, and correspondingly low rock/ice fraction, presents a puzzle when compared to the other three satellites, each of which is roughly $50\%$ ice and rock. In turn, the rock/ice fractions for Ganymede, Titan and Callisto are comparable to that of (captured) Saturnian irregular satellite Phoebe (or even smaller if Phoebe were porous).\\
Tying the properties of solar nebula planetesimals to subnebula satellitesimals involves three distinct aspects: first, characterising the properties of the first generation of planetesimals in terms of sizes and degree of heterogeneity; second, quantifying the collisional evolution of the planetesimal swarm following giant planet formation; and third, delivering planetesimal fragments to the circumplanetary disk. Planetesimal break-up in tandem with delivery via ablation of planetesimal fragments crossing the subdisk provides a framework for understanding the mass budget and compositions of regular satellites (compared to that of solar nebula planetesimals). In particular, ablation can result in fractionation, and account for the observed density of Iapetus provided this satellite formed in situ \citep{mae05}. For this to work planetesimals of size $\sim 10$ km need to be partially differentiated, which indicates that the first generation of planetesimals in the Jupiter-Saturn region (and possibly beyond) incorporated significant quantities of $^{26}$Al.\\ 
Thermal ablation occurs because friction of the fast moving heliocentric interloper as it crosses the circumplanetary gas disk heats up the body. At low gas densities impactors lose mass and energy through ablation. For large kilometer-sized bolides mechanical destruction rather than thermal ablation may dictate the fate of the object (see \citet{mec09} and references therein). Ablation of planetesimal fragments $< 1$ km may deliver the bulk of the solids needed to form the satellites. For objects in that size range, the heat transfer coefficient $C_H \sim 0.1$ may be reasonably obtained from observations of terrestrial meteorites \citep{bro83}. The rate at which energy is transferred to the planetesimal is given by $E \sim c_H \rho v^3$, where $\rho$ is the gas density, $v$ is the speed of the planetesimal through the gas, and some planetesimal flattening (which increases its cross section) takes place. One can then obtain an estimate of the surface temperature $T_s$ by balancing this heating and the radiative cooling. Ablation thus results in delivery of material to the circumplanetary disk. Gas drag may also result in capture of material.
To quantify these processes we first need to characterise the properties of the circumplanetary disk.\\
\begin{figure}
\centering
\includegraphics[width=\textwidth]{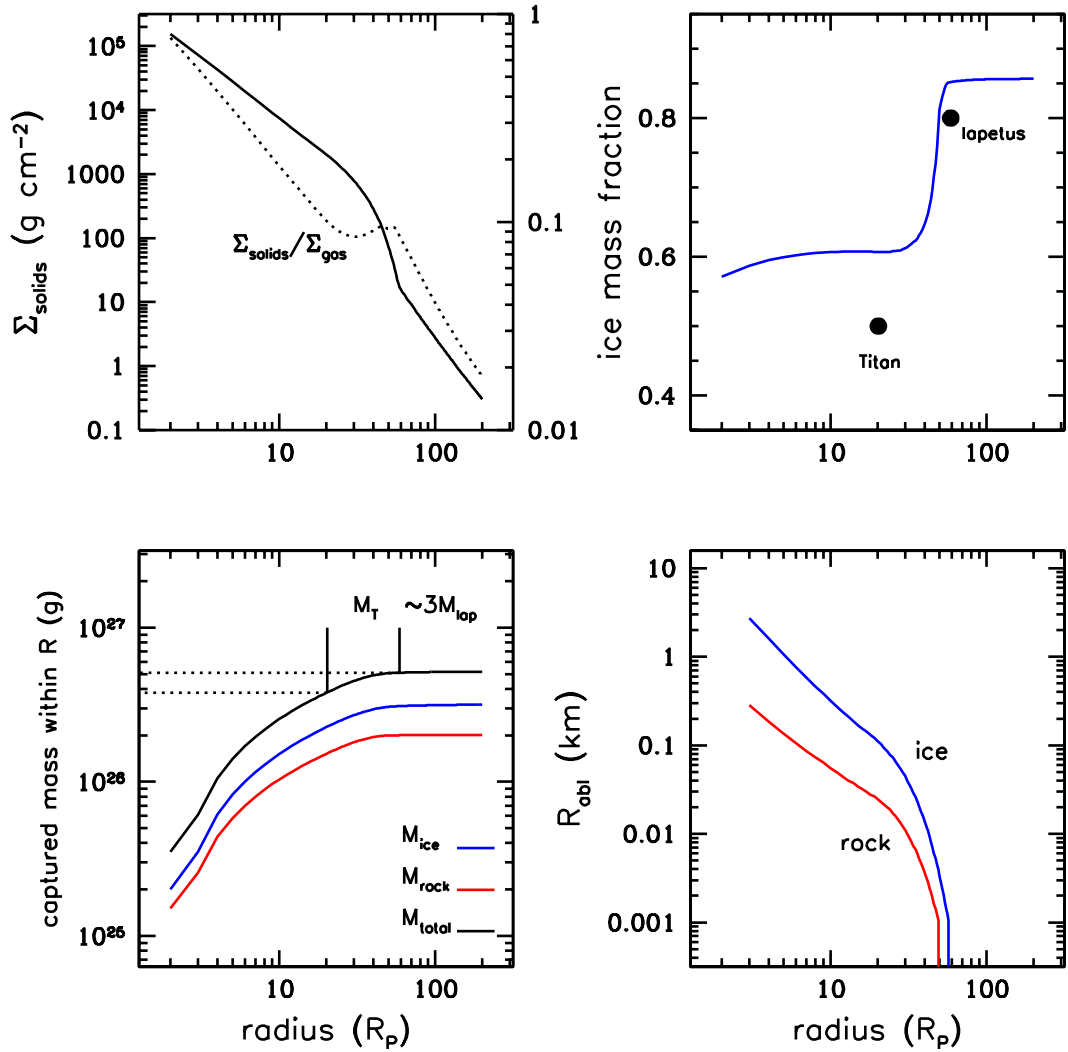}
\caption{Left upper panel: the solid-to-gas ratio (dotted line) and surface density of solids (solid line) as a function of radial distance from Saturn. Right upper panel: the ice to rock ratio as a function of radial distance. The empty dots correspond to Titan and Iapetus. The ice mass ratio of solar nebula planetesimal fragments crossing the disk is taken to be $0.3$. Left lower panel: the mass delivered within a given radius from Saturn. The lower right panel: the planetesimal size that gets either captured or ablated as function of radial distance.}\label{sec2fig1}
\end{figure}
In the SEMM model one expects a factor of $\sim 10$ enhancement in solids over cosmic mixtures, resulting in a gas surface density of $\sim 10^{4}$ g cm$^{-2}$, which is consistent with and quantitatively constrained by the gap-opening condition for Ganymede and Titan in a inviscid disk with aspect ratio $\sim 0.1$ \citep{raf02,mae03b}, and the Type I migration of full-sized satellites in such a disk \citep{mae03b,bea03}.
Following envelope collapse, planetesimal ablation of meter-to-kilometer disk crossers \citep{mae05} and other mechanisms enhanced the concentration of solids in the satellite disk. If we use a gas surface density of $10^4-10^5$ g cm$^{-2}$ for the Jovian and Saturnian subnebulae then a planetesimal of density $\sim 1$ g cm$^{-2}$ will encounter a gas column equal to its mass if its radius is in the range $0.1 - 1$ km. Following giant planet formation collisional fragmentation may replenish the $0.01 - 1$ km size range, even if most of the mass was originally in objects $> 1$ km. This means that a significant fraction of the mass in planetesimals crossing the gas disk may be ablated. Such a planetesimal may deposit a significant fraction of its mass in the gas disk.\\ 
In Figure \ref{sec2fig1} we show (see \citet{mec09} for further details) the mass delivery by ablation and gas drag capture of planetesimal fragments crossing the SEMM circumplanetary disk. These results are obtained assuming a differentiated population of icy and rocky interloper fragments with a rock fraction of $30\%$ by mass, and placing $\sim 10$ Earth masses in between Jupiter and Saturn. The bottom left panel shows that it is possible to deliver enough mass to account for the regular satellites of Jupiter and Saturn (only the Saturn results are shown). Furthermore, the ablation and the capture cross-section of meter-sized objects can lead to ice/rock fractionation and account for the composition of Iapetus, as well as those of Titan, Ganymede and Callisto (top right panel). This is the only explanation currently available for Iapetus' icy composition, and may also indicate that regular satellites overall may be depleted in rock with respect to solar composition mixtures, which might explain why their densities are lower than that of similar size Kuiper belt objects.\\ 
We stress that while full ice/rock separation is unlikely, only partial separation is needed to account for Iapetus' icy composition. Such a population can result from the fragmentation of a partially differentiated, $^{26}$Al-heated first generation of $\sim 10$ km planetesimals \citep{mec09}. Indeed, it is likely that most of the mass in the first generation of planetesimals resided in objects $10-100$ km in the first $10^5-10^6$ years, so that these objects may have incorporated significant amounts of $^{26}$Al. If so, $\sim 10-100$ km planetesimals can differentiate \citep{pap95} (Phoebe itself may be at least partially differentiated, see \citet{jea09}). A study by \citet{map06} that includes radioactive heating, accretion, transformation of amorphous to crystalline ice and melting of water ice in the formation of trans-neptunian objects (which take longer to grow than planetesimals forming closer in) with sizes $\sim 10$ km finds that the occurrence of liquid water may be common\footnote{But note that the presence of liquid water is not required for partial differentiation to take place. For instance, enhanced conductivity by water vapor flow through porous media may discourage melting, but would still result in a layered internal state as the core would become water-depleted \citep{pea08}.}. The disruption of partially differentiated planetesimals would then lead to a population of icy/rocky fragments available to ablate through the extended Kronian gas disk.

\section{Alien moons: the capture of Triton and the irregular satellites}

The origin and the subsequent evolution of the irregular satellites have profound implications for our understanding of the satellite systems of the giant planets. In this section we discuss the origins of Triton and of the irregular satellites. A separate chapter in this volume \citep{sea09} extends the discussion to the secular evolution of irregular satellites, with a focus on Phoebe.

\subsection{The irregular satellites}

Since the time of their discovery, irregular satellites have been known to belong to a separate population of objects. While the regular satellites orbit their parent planets on inner, equatorial and almost circular orbits, the orbits of the irregular satellites are about an order of magnitude larger, and are characterised by high eccentricities and inclinations. Moreover, a significant fraction of irregular satellites move on retrograde orbits (i.e., with inclinations in the range $[90^{\circ}-180^{\circ}]$). These dynamical features are incompatible with formation in prograde circumplanetary disks, which strongly suggests that irregular satellites are captured objects that originated elsewhere in the Solar System.\\
Depending on the formation regions of the parent bodies, irregular satellites would be characterised by different compositions. Therefore, once captured, irregular satellites would introduce exogenous materials into their host systems. Moreover, both theoretical studies and observational results suggest that mass transfer between regular and irregular satellites takes place, introducing contaminants on the surfaces of the inner moons. The identification of the irregular satellite capture mechanism (or mechanisms), and of the formation regions of the parent bodies, are thus key to our understanding of such contamination processes.

\subsubsection{Capture mechanisms and formation regions}

To permanently capture the parent bodies of the irregular satellites within the Hill radii of the giant planets, a fraction of their kinetic energy must be removed during passage. Historically, three dissipation mechanisms have been proposed to explain the capture of the irregular satellites: gas-drag \citep{pol79}, \emph{Pull-Down} \citep{hep77}, and collisional capture \citep{col71}. The first two scenarios are based on the presence of nebular gas and thus place the capture of the irregular satellites at the time of the formation of the giant planets. In the \citet{pol79} scenario gas is needed to slow the bodies crossing the Hill's spheres of the giant planets through friction. In the Pull-Down scenario by \citet{hep77} gas is needed to rapidly increase the mass of the giant planets and expand their Hill's spheres, thus trapping the irregular satellites in their enlarged gravitational fields. \citet{hep77} suggest that the gas-drag and Pull-Down mechanisms can be combined into a single scenario, where the interplay of the different processes can enhance the capture efficiency. We refer the reader to recent reviews (see e.g. \citet{jah07} and references therein) for more detailed discussions of the viability of gas-based scenarios; nevertheless, we point out that, while such scenarios can be applied to gas-rich Jupiter and Saturn, their applicability to Uranus and Neptune is uncertain. The collisional capture scenario by \citet{col71} has been somewhat neglected, yet it poses the least amount of constraints on the time of capture, has in principle the same capture efficiency when it comes to the gaseous and ice/rock giant planets, and may provide an explanation for collisional families among the irregular satellites of Jupiter. The Pull-Down scenario by \citet{hep77} explains the existence of these families by allowing for the post-capture disruption of more massive parent bodies through collisional events as postulated by \citet{col71}. The gas-drag model of \citet{pol79} accounts for the existence of collisional families through the break-up of captured bodies due to friction. However, owing to the size-dependence of gas drag, one might expect smaller bodies on inner orbits and larger ones on outer orbits, which is not observed in the present satellite systems.\\
More recently, a new class of dynamical models has been proposed to explain the capture of the irregular satellites. These new scenarios fit within the so-called \emph{Nice Model}, wherein it is argued that the structure of the outer Solar System is the result of a phase of chaotic rearrangement of the orbits of the giant planets moderated by the presence of a residual disk of planetesimals (for details we refer the reader to the original series of articles by \citet{gom05,mor05,tsi05}). Since the chaotic rearrangement of the orbits of the giant planets would destroy any pre-existing system of irregular satellites (see \citet{nes07}, and references within), gas-based models would not be viable, with Jupiter as the only possible exception due to its limited participation in the orbital rearrangement. Proposed mechanisms to resupply the populations of irregular satellites rely on exchange reactions (i.e., the disruption of a binary system of planetesimals while crossing the Hill's sphere of a giant planet, see \citet{agn06,vok08}), and three-body effects due to the mutual gravitational perturbations of the giant planets during close encounters in a Nice Model-like scenario \citep{nes07}. In addition, collisional capture has been studied in more detail for Saturn's system in light of the observational data supplied by Cassini-Huygens mission \citep{tur09}.\\
A detailed discussion of capture mechanisms is beyond the scope of this chapter. In the framework of the processes shaping the moons of the giant planets, the most important influence of the irregular satellites is that of contaminating the surfaces of the inner satellites, introducing exogenous elements with potentially different compositional features. The nature of the contaminants delivered depends on the composition of the irregular satellites, which is linked to the formation regions of the parent bodies. The underlying assumption of early studies is that the source of the parent bodies of the irregular satellites is located near the orbital regions of the host planets. This would imply that one should expect a compositional gradient with heliocentric distance between the families of irregular satellites of the different planets, and that the overall composition of the irregular satellites should be compatible with that of the regular ones. To the contrary, the investigation of the collisional capture scenario performed by \citet{tur09} shows that the possible orbits of the putative parent bodies of the irregular satellites span over a wide range of semimajor axes and eccentricity values, covering the whole outer Solar System. This means that irregular satellites orbiting a given planet could have formed in different regions of the Solar System, and thus have significant differences in their compositions. The same conclusion also holds true in the Nice Model, due to the orbital rearrangement suffered by the early Solar System.

\subsubsection{Implications of the capture origins}

While our data on the irregular satellites of Uranus and Neptune are still limited, both dynamical and observational evidence suggests that collisions between members of the populations of irregular satellites of Jupiter and Saturn played an important role in the past. The orbital structure of the Jovian irregular satellites bears a clear imprint of the existence of collisional families \citep{nes03,nes04}. The orbital structure of the Kronian irregular satellites is more difficult to interpret, due to Jupiter's strong gravitational perturbations, yet there are clues of the possible existence of collisional families \citep{tur08}. Moreover, there are strong dynamical indications that Phoebe underwent a period of intense collisional activity in the past with now-extinct irregular satellites located near its orbital region \citep{tur08}, and such a hypothesis is supported by the observations of the instruments on-board Cassini of Phoebe's deeply cratered surface \citep{por05}. The collisional activity of the irregular satellites would produce fragments, which could be ejected on orbits that intersect those of the regular satellites. Depending on their size, dust grains would be expelled into heliocentric orbits due to radiation pressure, or migrate inward due to Poynting-Robertson drag. Dust migrating inward would intersect the orbits of the regular satellites and, depending on their sweeping efficiency, reach different depths in the satellite systems. Not all dust grains are collected with the same efficiency: bigger grains migrate more slowly, enhancing the probability for the inner satellites to capture them. Also, dust particles in retrograde orbits would experience more close encounters with the satellites, and are more likely to be collected than grains in prograde orbits. Since the parent bodies of the irregular satellites originated in very different environments, this dust production/transfer leads to the contamination of the surfaces of the icy satellites with exogenous material characterised by a different chemical composition than the local material, and possibly coming from different sources. The origin of the dark material coating the leading hemisphere of Iapetus is probably the best example of this process. Still, whether the cause of the leading-trailing side dichotomy on Iapetus is endogenous, as opposed to exogenous, continues to be debated.

\subsection{The case of Triton}

Another case in point is Triton, where the transfer process acts in the opposite way, i.e., the regular satellites are the source of contamination on Triton. Like other irregular satellites, Triton's retrograde orbit points to a capture origin \citep{gea89,agn06}. Yet, Triton's semimajor axis is over an order of magnitude smaller than that of the other irregular satellites, and Triton's size makes it one of the most massive moons in the Solar System. The small orbital distance of the satellite from Neptune suggests that its close-in, circular orbit is the result of tidal dissipation \citep{gea89}; it has been pointed out, however, that Triton's high inclination ($i\approx157^{\circ}$) implies a time-scale of the order of the age of the Solar System for tidal dissipation to shrink the orbit of the satellite from probable post-capture values of the semimajor axis to its present location \citep{cag05}. Triton's gravitational perturbations from its initial post-capture orbit would have caused pre-existing regular satellites to collide with each other and shatter. Taking the mass of the pre-existing satellites to be comparable to that of Triton, which these authors justify by analogy with the Uranian satellite system, the resulting debris disk could have shrunk Triton's orbit through scatterings and collisional debris-drag on a time-scale of about $10^{5}$ years, possibly allowing for the survival of the outermost regular and irregular satellites. Such a hypothesis might be in agreement with the results obtained by \citet{saz07} in reprocessing the Voyager's data on Triton, suggesting that the asymmetric distribution of craters on the surface of the satellite could be best explained by assuming a planetocentric source for the impactors, i.e., that the craters on Triton's leading hemisphere are due to impacts with a population of prograde small satellites or collisional shards. The paucity of craters on Triton could be an indication that the surface of the satellite is extremely young, with its age varying in the range $10^{7}-10^{8}$ years, diagnostic of an active geology possibly linked to the tidal evolution of the satellite. This young surface age could indicate that the planetocentric impactors responsible for the craters could be the surviving members of the original debris disk. While the possibility of linking Triton's origin and evolution to its present state is intriguing, it should be noted that at the moment no quantitative study of the collisional lifetime of such a debris disk exists. Moreover, the data available on Triton and on Neptune's system in general are quite limited, making it difficult to derive better observational constrains for the theoretical models. Nevertheless, it is clear that one needs to take into account Triton's history and its exogenous origin to properly interpret the data currently available and those to be supplied by future space missions. While the Voyager's data indicate that Triton's surface is quite young and possibly uncontaminated by external material, the craters on the leading hemisphere could contain residual material from the impactors, perhaps more similar in composition to the present regular satellites of Neptune than to the uncontaminated surface of Triton. At any rate, Triton's capture disrupted Neptune's pre-existing satellite system, implying that care needs to be exercised when attempting to extract information about proto-Neptune's subnebula from the structure of the present satellite system of the giant planet.

\section{Evolution: the path to present}

As planetary formation sets the boundary conditions for satellite formation, so satellite formation sets the initial conditions for satellite evolution. Dynamical and geochemical evolution, in fact, secularly reshapes the satellites, erasing or modifying their primordial features. Therefore, in order to enhance our understanding of the present state of the moons of the outer Solar System, we need to be able to discriminate the effects of their secular evolution from those properties inherited from their time of formation. Nevertheless, thanks to the recent Cassini mission to Saturn and the efforts of those who made it possible, we currently have in hand a sufficient number of reliable observational constraints to give us justifiable hope that it is indeed possible to make significant progress in our understanding of the origin and evolution of the satellite systems of the giant planets.

\end{document}